\documentclass[journal]{vgtc}                
\ifpdf
  \pdfoutput=1\relax                   
  \usepackage{graphicx}                
  \DeclareGraphicsExtensions{.pdf,.png,.jpg,.jpeg} 
\else
  \usepackage{graphicx}                
  \DeclareGraphicsExtensions{.eps}     
\fi%

\graphicspath{{figures/}{pictures/}{images/}{./}} 

\usepackage{microtype}                 
\PassOptionsToPackage{warn}{textcomp}  
\usepackage{textcomp}                  
\usepackage{mathptmx}                  
\usepackage{times}                     
\usepackage{cite}                      
\usepackage{tabu}                      
\usepackage{booktabs}                  

\usepackage{enumitem}
\setitemize{noitemsep,topsep=5pt,parsep=0pt,partopsep=0pt}
\usepackage{xspace}
\usepackage{xstring}
\usepackage[normalem]{ulem} 
\usepackage{amsmath}

\newcommand{\breakline}{\vskip \medskipamount}
\newcommand{\q}[1]{\noindent \textit{#1}}
\newcommand{\fakeTitle}[1]{\breakline \noindent \textbf{#1}.}

\newcommand{\vv}[1]{#1}

\newcommand{\ie}{{\em i.e.,}\xspace}
\newcommand{\eg}{{\em e.g.,}\xspace}

\newcommand{\etal}{{\em et~al.}\xspace}
\newcommand{\aka}{{\em a.k.a.}\xspace}

\newcommand{\figureTeaser}{
\teaser{
  \centering
  	\includegraphics[width=\linewidth,trim={0cm 13.9cm 3cm 0cm}, clip]{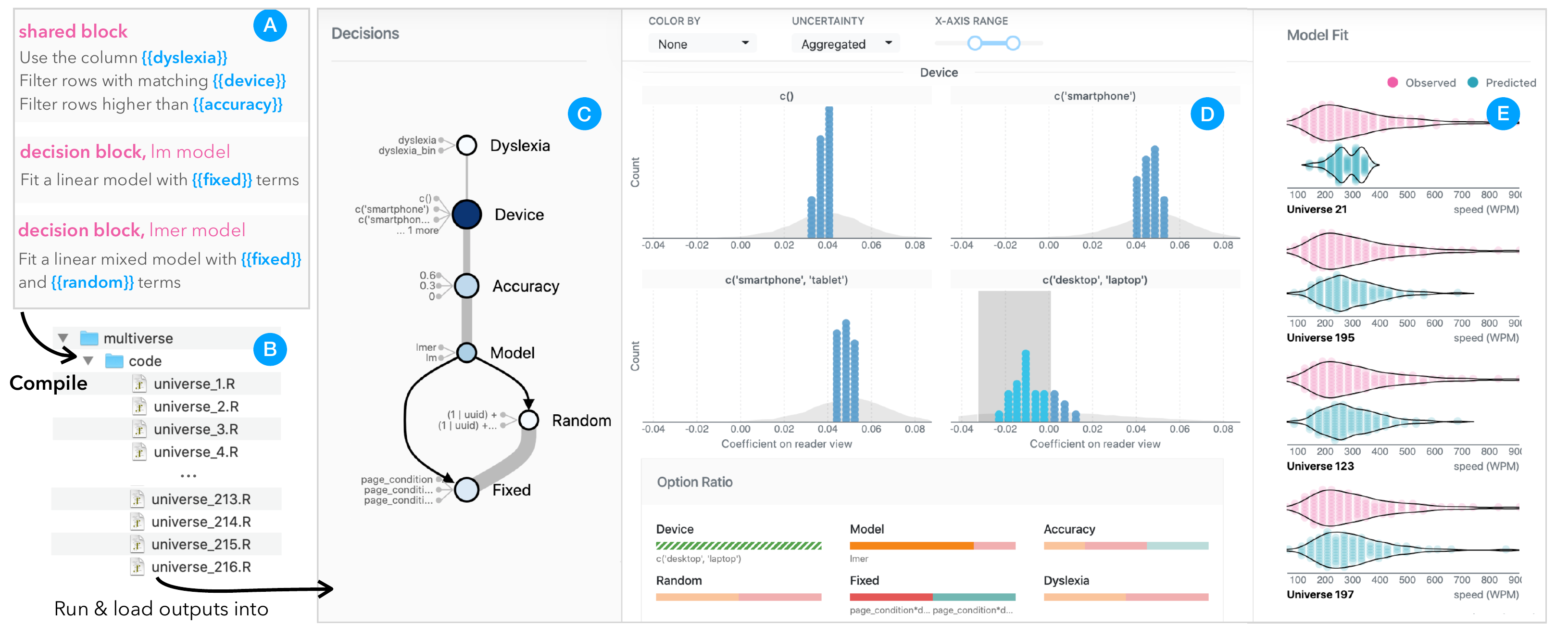}
  \vspace{-16pt}
  \caption{Authoring and visualizing multiverse analyses with Boba. Users start by annotating a script with analytic decisions (a), from which Boba synthesizes a multiplex of possible analysis variants (b). To interpret the results from all analyses, users start with a graph of analytic decisions (c), where sensitive decisions are highlighted in darker blues. Clicking a decision node allows users to compare point estimates (d, blue dots) and uncertainty distributions (d, gray area) between different alternatives. Users may further drill down to assess the fit quality of individual models (e) by comparing observed data (pink) with model predictions (teal).}
	\label{fig:teaser}
}
}

\newcommand{\figureWorkflow}{
\begin{figure}[t]
	\centering
	\includegraphics[width=0.8\columnwidth, trim={0cm 21.5cm 16cm 0cm}, clip]{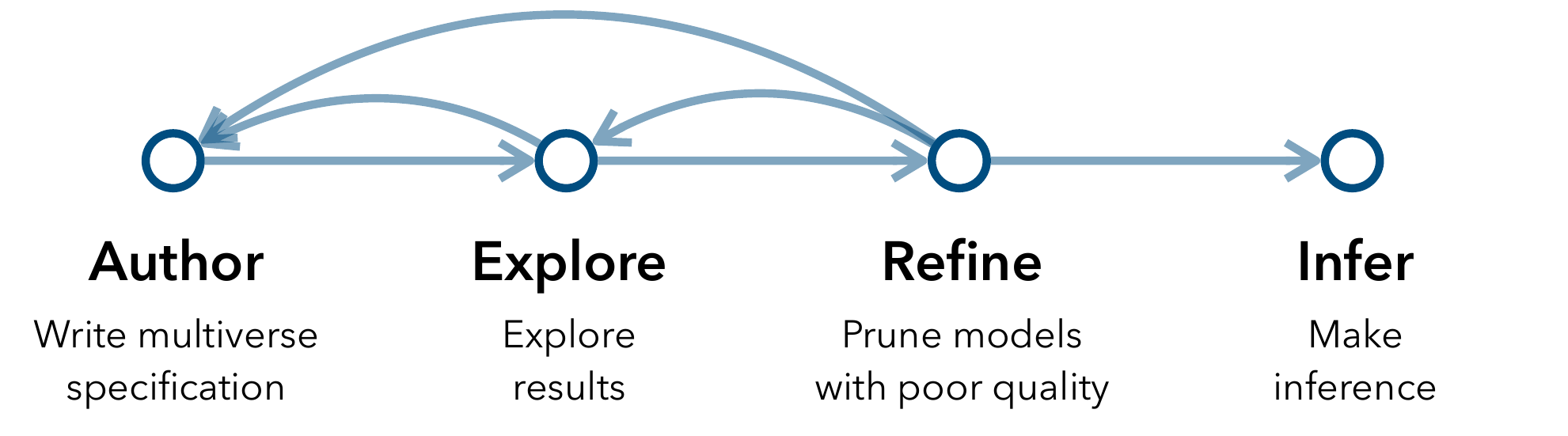}
	\vspace{-12pt}
    \caption{The intended workflow for multiverse analysis in Boba.}
    \label{fig:workflow}
    \vspace{-16pt}
\end{figure}
}

\newcommand{\figureSyntax}{
\begin{figure}[t]
    \centering
	\includegraphics[width=0.9\columnwidth,trim={1.5cm 5cm 6.5cm 1cm}, clip]{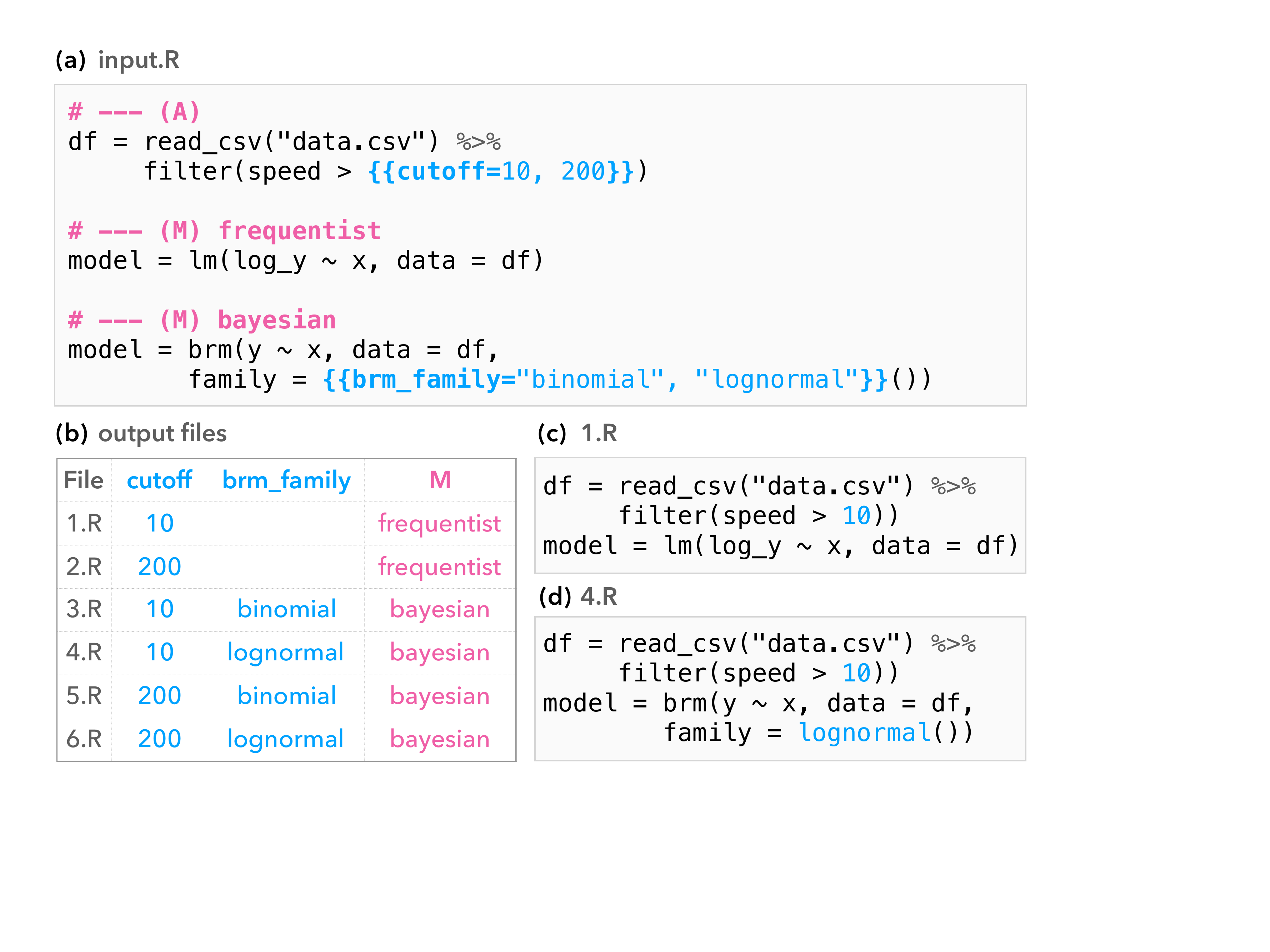}
	\vspace{-10pt}
    \caption{An example Boba specification. The user annotates an R script~(a) with two placeholder variables (blue) and three code blocks (pink). The compiler synthesizes six files (b). In the example output files (c) and (d), placeholder variables are replaced by their possible values, and only one version of the decision block M is present.}
    \label{fig:syntax}
    \vspace{-16pt}
\end{figure}
}

\newcommand{\figureDslExample}{
\begin{figure}[t]
    \begin{minipage}[t]{0.407\columnwidth}
        \vspace{0pt}
        \includegraphics[width=1.05\columnwidth,trim={0cm 6.8cm 20cm 0cm}, clip]{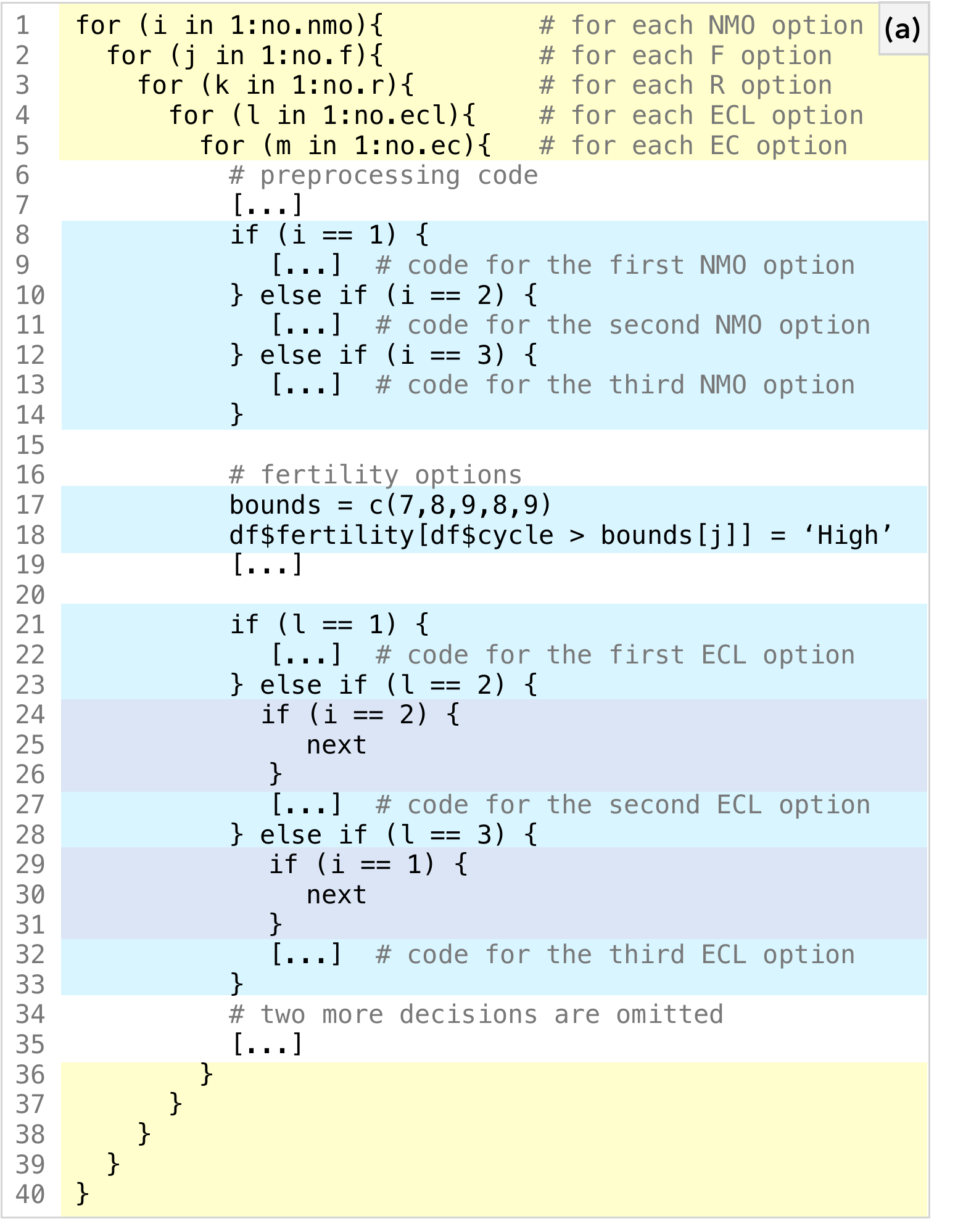}
    \end{minipage}
    \begin{minipage}[t]{0.591\columnwidth}
        \vspace{0pt}
        \includegraphics[width=1.05\columnwidth,trim={0cm 13cm 20cm 0cm}, clip]{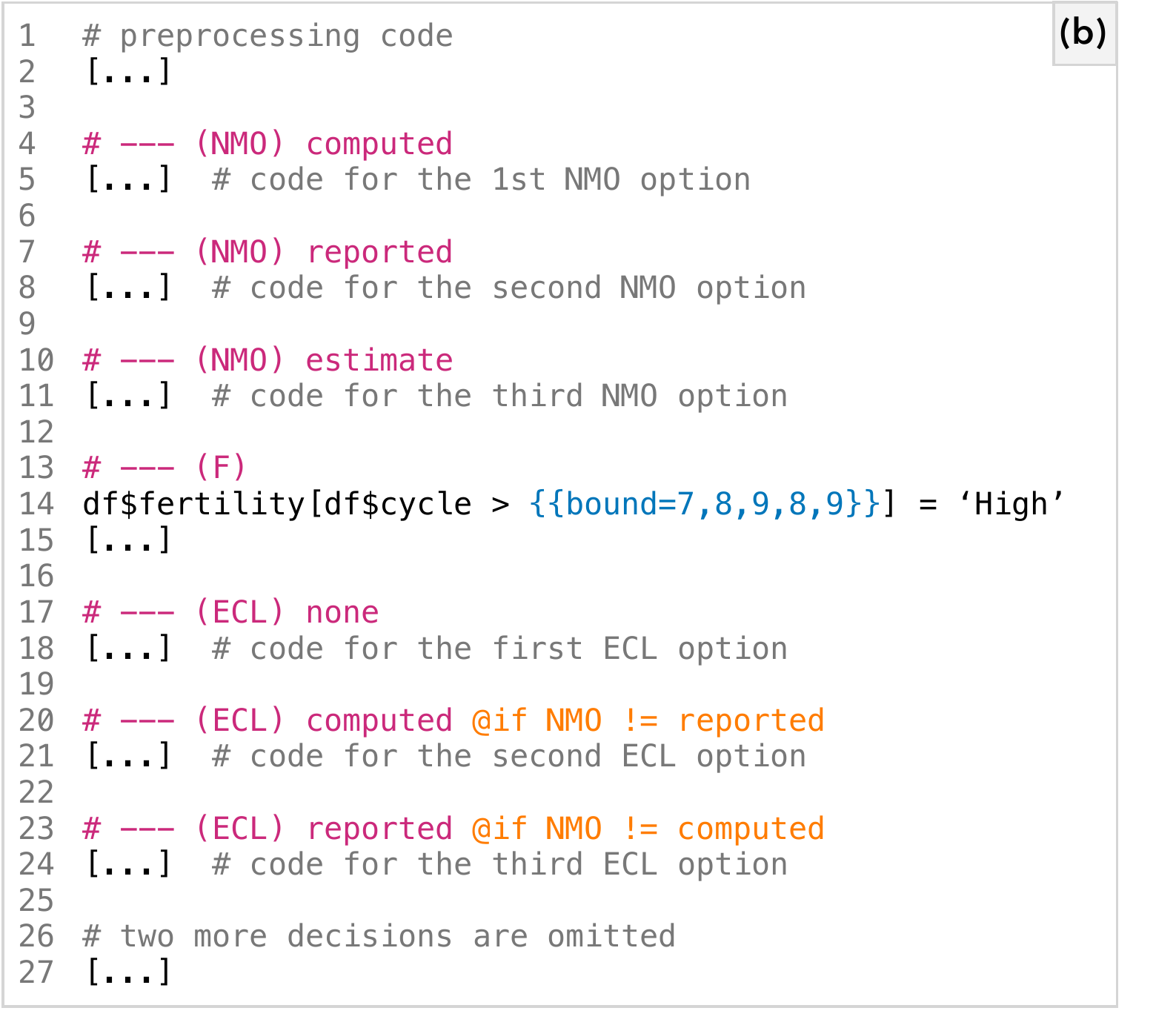}
    \end{minipage}
	\vspace{-10pt}
    \caption{Specification of a real-world multiverse analysis~\cite{steegen2016} with five decisions and a procedural dependency. (a) Markup of the R code written by original authors, with custom control flow (nested for-loops and if-statements) highlighted. (b) Markup of the Boba DSL specification.}
    \label{fig:dsl-example}
    \vspace{-10pt}
\end{figure}
}

\newcommand{\figureSystemStartup}{
\begin{figure}[t]
    \centering
	\includegraphics[width=1.0\columnwidth,trim={0cm 16cm 15.5cm 0cm}, clip]{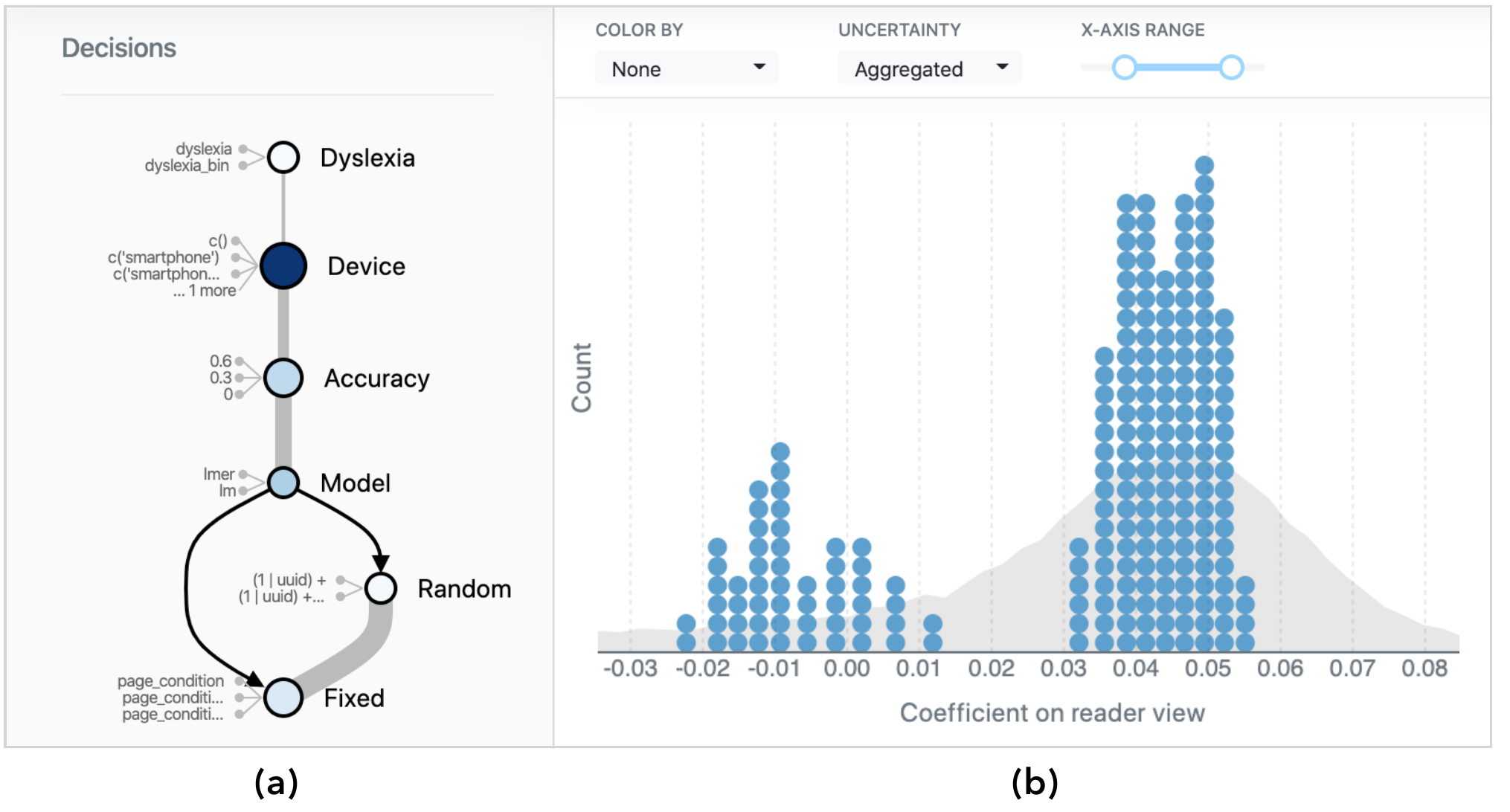}
	\vspace{-22pt}
    \caption{\textbf{Decision view and outcome view}. (a) The decision view shows analytic decisions as a graph with order and dependencies between them, and highlights more sensitive decisions in darker colors. (b) The outcome view visualizes outputs from all analyses, including individual point estimates and aggregated uncertainty.}
    \label{fig:system-startup}
    \vspace{-16pt}
\end{figure}
}

\newcommand{\figureFacetBrush}{
\begin{figure}[t]
    \centering
	\includegraphics[width=1.0\columnwidth,trim={0cm 10cm 8.5cm 0cm}, clip]{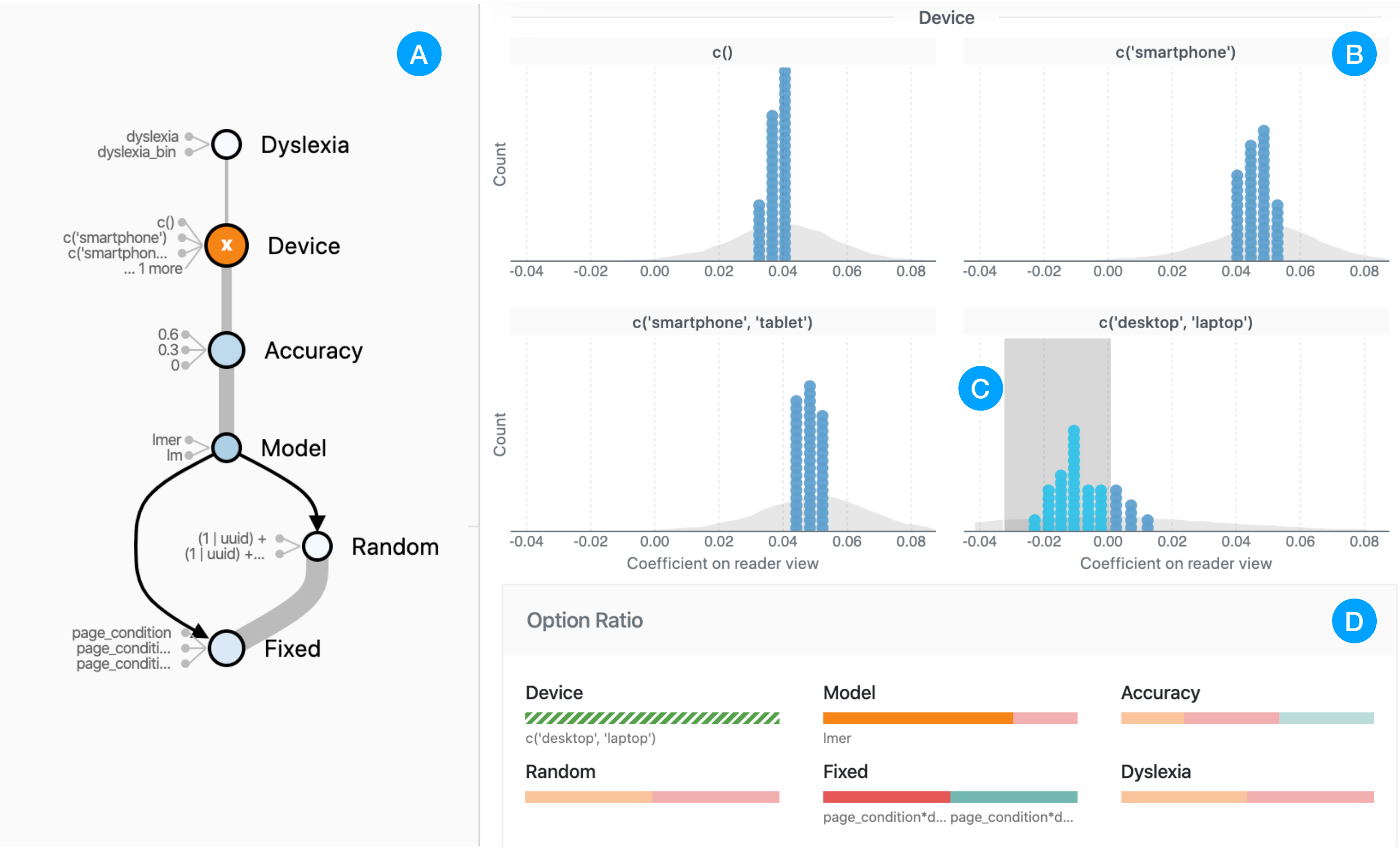}
	\vspace{-22pt} 
    \caption{\textbf{Facet and Brushing}. Clicking a node in the decision view (a) divides the outcome view into a trellis plot (b), answering questions like ``does the decision lead to large variations in effect size?'' Brushing a region in the outcome view (c) reveals dominant alternatives in the option ratio view (d), answering questions like ``what causes negative results?''}
    \label{fig:system-facet}
    \vspace{-8pt} 
\end{figure}
}

\newcommand{\figureCurves}{
\begin{figure}[t]
    \centering
	\includegraphics[width=1.0\columnwidth,trim={0cm 15.7cm 13.5cm 0cm}, clip]{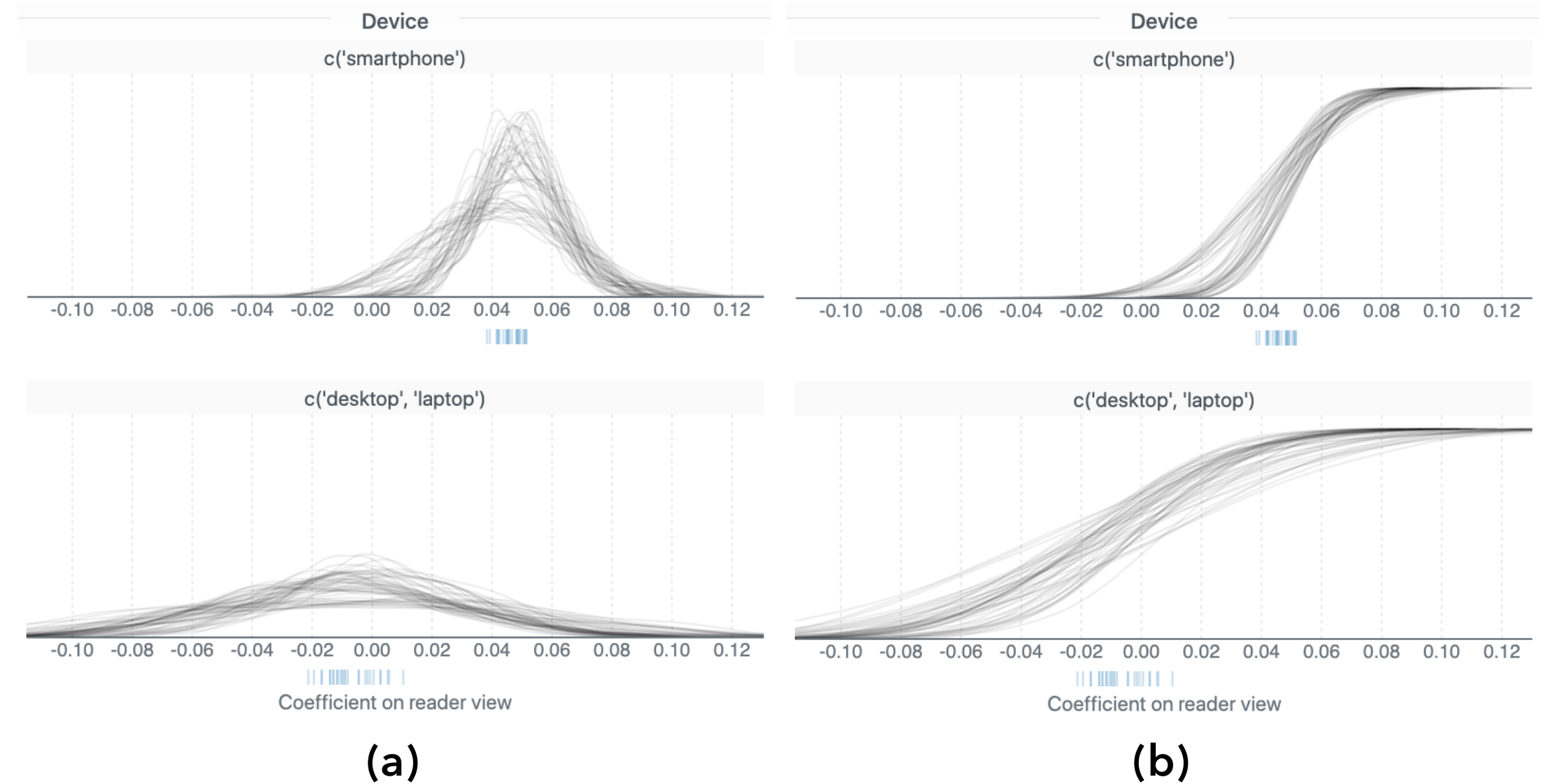}
	\vspace{-20pt} 
    \caption{\textbf{PDFs} (a) and \textbf{CDFs} (b) \textbf{views} visualize sampling distributions from individual universes. Toggling these views in a trellis plot allows users to compare the variance between conditions.}
    \label{fig:system-curves}
    \vspace{-16pt}
\end{figure}
}

\newcommand{\figurePrune}{
\begin{figure}[t]
    \centering
	\includegraphics[width=0.98\columnwidth, trim={0cm 17cm 9cm 0cm}, clip]{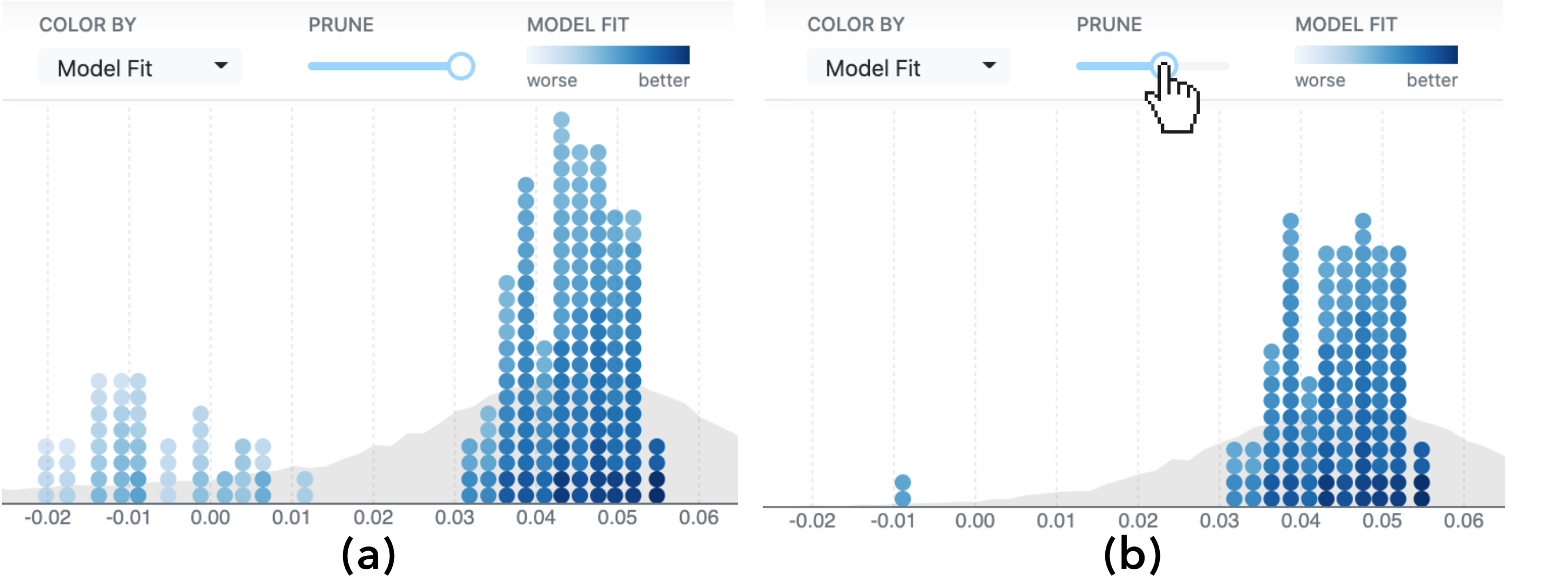}
	\vspace{-12pt} 
    \caption{(a) Coloring the universes according to their model fit quality. (b) Removing universes that fail to meet a model quality threshold.}
    \label{fig:system-prune}
    \vspace{-5pt}
\end{figure}
}

\newcommand{\figureInference}{
\begin{figure}[t]
    \centering
	\includegraphics[width=0.8\columnwidth, trim={0cm 18.5cm 23.4cm 0cm}, clip]{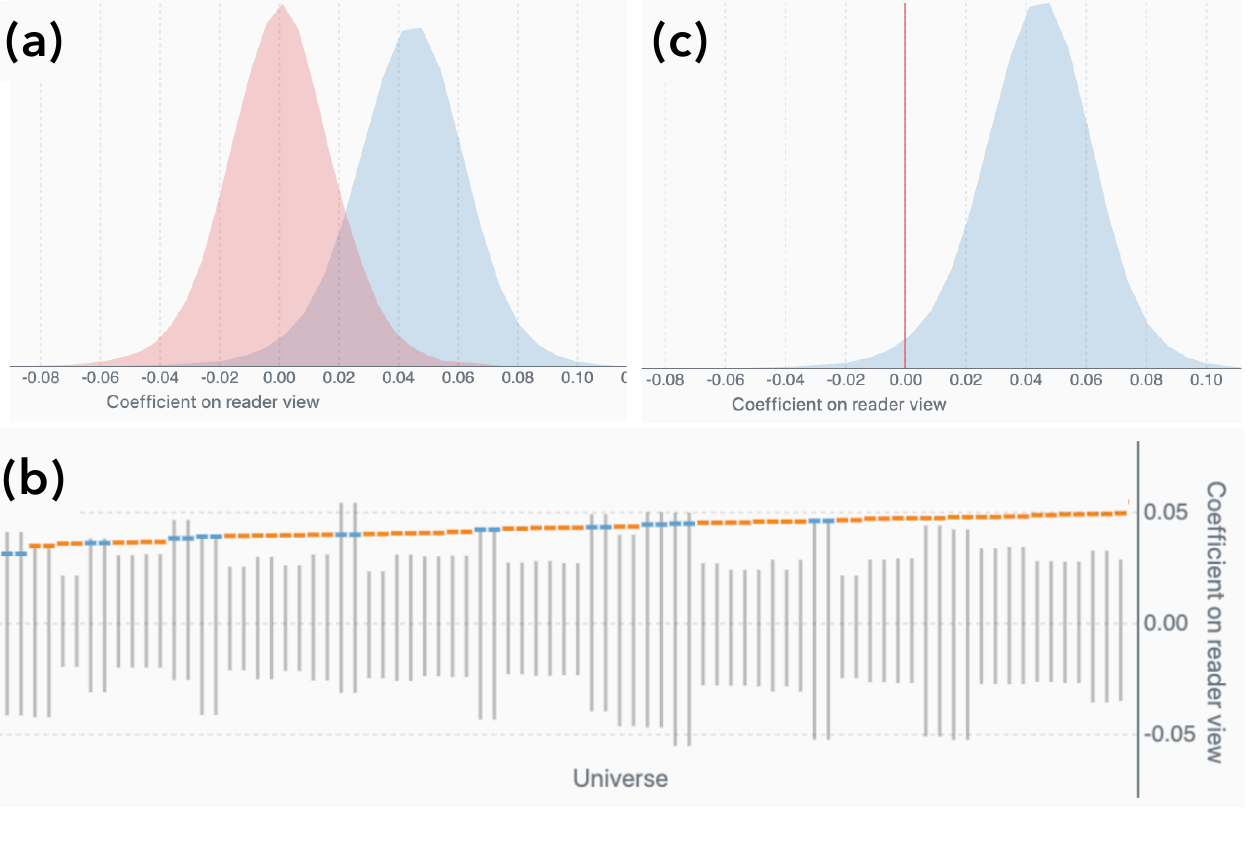}
	\vspace{-12pt} 
    \caption{\textbf{Inference views.} (a) Aggregate plot comparing the possible outcomes of the actual multiverse (blue) and the null distribution (red). (b) Detailed plot showing the individual point estimates and the range between the 2.5th and 97.5th percentile in the null distribution (gray line). Point estimates outside the range are colored in orange. (c) Alternative aggregate plot where a red line marks the expected null effect. }
    \label{fig:system-inference}
    \vspace{-12pt}
\end{figure}
}

\newcommand{\figureMortgage}{
\begin{figure*}[t]
    \centering
	\includegraphics[width=1.9\columnwidth, trim={0cm 10cm 0cm 0cm}, clip]{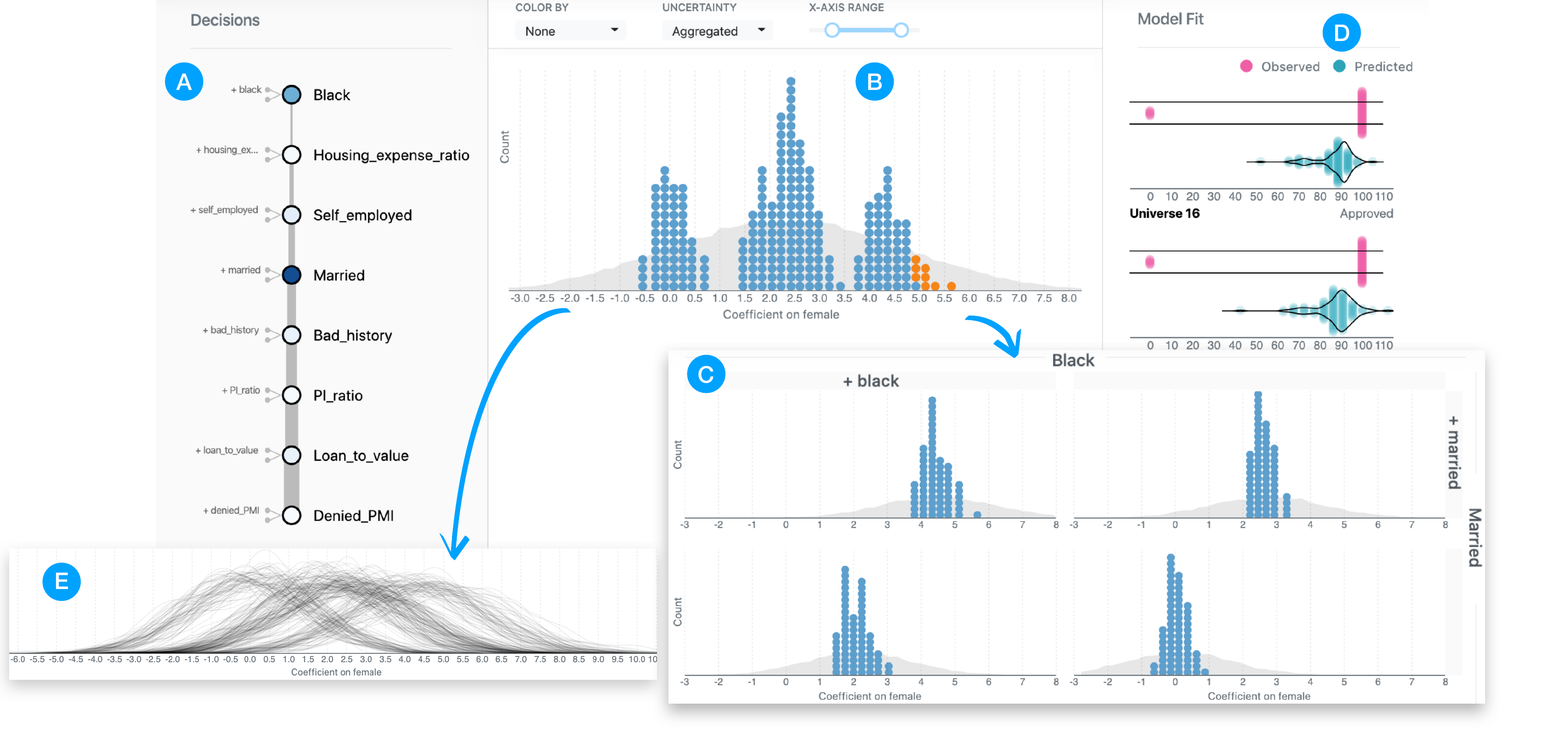}
	\vspace{-18pt} 
    \caption{A case study on how model estimates are robust to control variables in a mortgage lending dataset. (a) Decision view shows that \textit{black} and \textit{married} are two consequential decisions. (b) Overall outcome distribution follows a multimodal distribution with three peaks. (c) Trellis plot of \textit{black} and \textit{married} indicates the source of the peaks. (d) Model fit plots show that models produce numeric predictions while observed data is categorical. (e) PDFs of individual sampling distributions show significant overlap of the three peaks.}
    \label{fig:mortgage}
    \vspace{-14pt}
\end{figure*}
}

\newcommand{\figureHurricane}{
\begin{figure*}[t]
    \centering
	\includegraphics[width=1.6\columnwidth, trim={0cm 13cm 14cm 0cm}, clip]{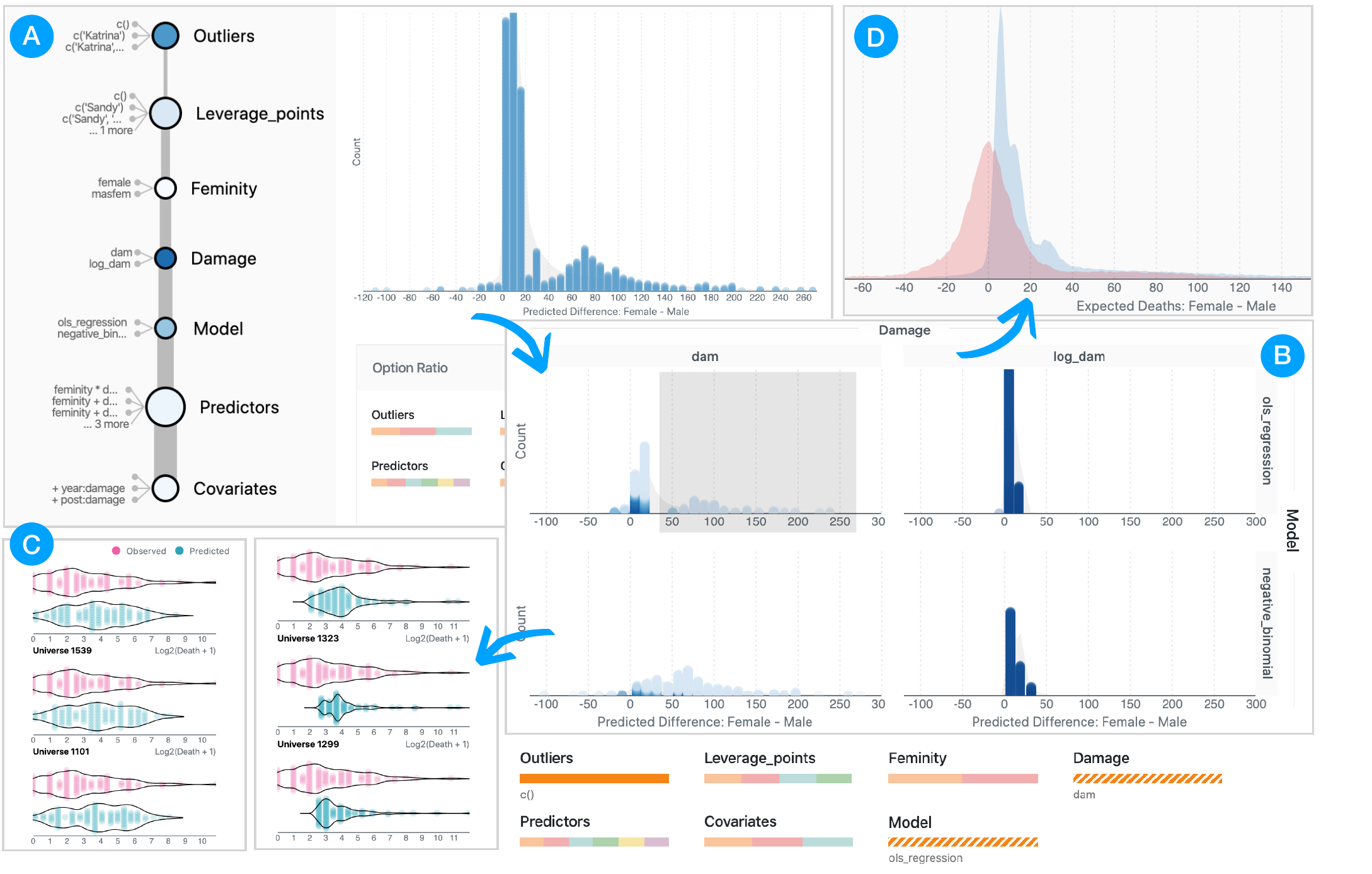}
	\vspace{-12pt} 
    \caption{A case study on whether hurricanes with more feminine names have caused more deaths. (a) The majority of point estimates suggest a small, positive effect, but there are considerable variations. (b) Faceting and brushing reveal decision combinations that produce large estimates. Coloring by model quality shows that large estimates are from questionable models, and predictive checks (c) confirms model fit issues. (d) Inference view shows that the observed and null distributions are different in terms of mode and shape, yet with highly overlapping estimates.}
    \label{fig:hurricane}
    \vspace{-10pt}
\end{figure*}
}

\vgtccategory{Research}
\vgtcpapertype{system}

\title{Boba: Authoring and Visualizing Multiverse Analyses}

\author{Yang Liu, Alex Kale, Tim Althoff, and Jeffrey Heer}
\authorfooter{
\item
 All authors are with the University of Washington. E-mails: yliu0, kalea, althoff, jheer@uw.edu.
}

\shortauthortitle{Liu \MakeLowercase{\textit{et al.}}: Boba: Authoring and Visualizing Multiverse Analyses}

\abstract{Multiverse analysis is an approach to data analysis in which all ``reasonable'' analytic decisions are evaluated in parallel and interpreted collectively, in order to foster robustness and transparency.
However, specifying a multiverse is demanding because analysts must manage myriad variants from a cross-product of analytic decisions, and the results require nuanced interpretation.
We contribute Boba: an integrated domain-specific language (DSL) and visual analysis system for authoring and reviewing multiverse analyses.
With the Boba DSL, analysts write the shared portion of analysis code only once, alongside local variations defining alternative decisions, from which the compiler generates a multiplex of scripts representing all possible analysis paths.
The Boba Visualizer provides linked views of model results and the multiverse decision space to enable rapid, systematic assessment of consequential decisions and robustness, including sampling uncertainty and model fit.
We demonstrate Boba's utility through two data analysis case studies, and reflect on challenges and design opportunities for multiverse analysis software. 
}


\CCScatlist{ 
 \CCScat{K.6.1}{Management of Computing and Information Systems}%
{Project and People Management}{Life Cycle};
 \CCScat{K.7.m}{The Computing Profession}{Miscellaneous}{Ethics}
}

\figureTeaser

\vgtcinsertpkg

\begin{document}



\maketitle

\section{Introduction} 

The last decade saw widespread failure to replicate findings in published literature across multiple scientific fields~\cite{begley2012, border2019, open2015, prinz2011}.
As the replication crisis emerged~\cite{baker2016}, scholars began to re-examine how data analysis practices might lead to spurious findings.
An important contributing factor is the flexibility in making analytic decisions~\cite{simmons2011, gelman2013garden, gelman2014}.
Drawing inferences from data often involves many decisions: what are the cutoffs for outliers? What covariates should one include in statistical models?
Different combinations of choices might lead to diverging results and conflicting conclusions. 
Flexibility in making decisions might inflate false-positive rates when researchers explore multiple alternatives and selectively report desired outcomes~\cite{simmons2011}, a practice known as p-hacking~\cite{nelson2018}.
Even without exploring multiple paths, fixating on a single analytic path might be less rigorous, as multiple justifiable alternatives might exist and picking one would be arbitrary. 
For example, a crowdsourced study~\cite{silberzahn2018} shows that well-intentioned experts still produce large variations in analysis outcomes when analyzing the same dataset independently.

In response, prior work proposes \textit{multiverse analysis}, an approach to outline all ``reasonable'' alternatives a-priori, exhaust all possible combinations between them, execute the end-to-end analysis per combination, and interpret the outcomes collectively~\cite{simonsohn2015, steegen2016}.
A multiverse analysis demonstrates the extent to which conclusions are robust to sometimes arbitrary analytic decisions.
Furthermore, reporting the full range of possible results, not just those which fit a particular hypothesis or narrative, helps increase the transparency of a study~\cite{rubin2017}.

However, researchers face a series of barriers when performing multiverse analyses.
Authoring a multiverse is tedious, as researchers are no longer dealing with a single analysis, but hundreds of forking paths resulting from possible combinations of analytic decisions.
Without proper scaffolding, researchers might resort to multiple, largely redundant analysis scripts~\cite{kery2017}, or rely on \vv{intricate control flow structure} including nested for-loops and if-statements~\cite{steegen_osf}.
Interpreting the outcomes of a vast number of analyses is also challenging.
Besides gauging the overall robustness of the findings, researchers often seek to understand what decisions are critical in obtaining particular outcomes (\eg ~\cite{simonsohn2015, steegen2016, young2017}).
As multiple decisions might interact, understanding the nuances in how decisions affect robustness will require a comprehensive exploration, suggesting a need for an interactive interface.


To lower these barriers, we present Boba, an integrated domain-specific language (DSL) and visualization system for multiverse authoring and interpretation. 
Rather than managing myriad analysis versions in parallel, the Boba DSL allows users to specify the shared portion of the analysis code only once, alongside local variations defining alternative analysis decisions.
The compiler enumerates all compatible combinations of decisions and synthesizes individual analysis scripts for each path.
As a meta-language, the Boba DSL is agnostic to the underlying programming language of the analysis script (\eg Python or R), thereby supporting a wide range of data science use cases.

The Boba Visualizer facilitates assessment of the output of all analysis paths. 
We support a workflow where users view the results, refine the analysis based on model quality, and commit the final choices to making inference.
The system first provides linked views of both analysis results and the multiverse decision space to enable a systematic exploration of how decisions do (or do not) impact outcomes.
Besides decision sensitivity, we enable users to take into account sampling uncertainty and model fit by comparing observed data with model predictions~\cite{Gelman2003}. 
After viewing the results, users can exclude models poorly suited for inference by adjusting a model fit threshold, or adopt a principled approach based on model averaging to incorporate model fit in inference.
We discuss the implications of post-hoc refinement, along with other challenges in multiverse analysis, in our design reflections.


We evaluate Boba in a code comparison example and two data analysis case studies.
We first demonstrate how the Boba DSL eliminates custom control-flows when implementing a real-world multiverse of considerable complexity.
Then, in two multiverses replicated from prior work~\cite{simonsohn2015, young2017}, we show how the Boba Visualizer affords multiverse interpretation, enabling a richer understanding of robustness, decision patterns, and model fit quality via visual inspection.
In both case studies, model fit visualizations surface previously overlooked issues and change what one can reasonably take away from these multiverses.


\section{Related Work}\label{sec:related-work}

We draw on prior work on authoring and visualizing multiverse analyses, and approaches for authoring alternative programs and designs.

\subsection{Multiverse Analysis}

Analysts begin a multiverse analysis by identifying reasonable analytic decisions \textit{a-priori}~\cite{simonsohn2015,steegen2016, patel2015}. 
Prior work defines reasonable decisions as those with firm theoretical and statistical support~\cite{simonsohn2015}, and decisions can span the entire analysis pipeline from data collection and wrangling to statistical modeling and inference~\cite{wicherts2016, liu2020-paths}.
While general guidelines such as a decision checklist~\cite{wicherts2016} exist, defining what decisions are reasonable still involves a high degree of researcher subjectivity.

The next step in multiverse analyses is to exhaust all compatible decision combinations and execute the analysis variants (we call a variant a \textit{universe}).
Despite the growing interest in performing multiverse analysis (\eg~\cite{jelveh2018, orben2019, crede2017, border2019, rohrer2017}), few tools currently exist to aid authoring.
Young and Holsteen~\cite{young2017} developed a STATA module that simplifies multimodel analysis into a single command, but it only works for simple variable substitution. 
\textit{Rdfanalysis}~\cite{gassen2019}, an R package, supports more complex alternative scenarios beyond simple value substitution, but the architecture assumes a linear sequential relationship between decisions.
Our DSL similarly provides scaffolding for specifying a multiverse, but it has a simpler syntax, extends to other languages, and handles procedural dependencies between decisions.

After running all universes,
the next task is to interpret results collectively. 
Some prior studies visualize results from individual universes by either juxtaposition~\cite{simonsohn2015, steegen2016, rae2019} or animation~\cite{dragicevic2019}.
Visualizations in other studies apply aggregation~\cite{dejonckheere2018, poarch2019}, for example showing a histogram of effect sizes.
The primary issue with juxtaposing or animating individual outcomes is scalability, 
though this might be circumvented by sampling~\cite{simonsohn2015}. 
Our visualizer shows individual outcomes, but overlays or aggregates outcomes in larger multiverses to provide scalability.

Besides the overall robustness, many studies also investigate which analytic decisions are most consequential.
The simplest approach is a table~\cite{steegen2016, cesario2019, rae2019, dejonckheere2019} where rows and columns map to decisions, and cells represents outcomes from individual universes.
Simonsohn et al.~\cite{simonsohn2015} extend this idea, visualizing the decision space as a matrix beneath a plot of sorted effect sizes.
These solutions might not scale as they juxtapose individual outcomes, and the patterns of how outcomes vary might be difficult to identify depending on the spatial arrangements of rows and columns.
Another approach~\cite{poarch2019} slices the aggregated distribution of outcomes along a decision dimension to create a trellis plot (\aka small multiples~\cite{tufte1990}).
The trellis plot shows how results vary given a decision, but does not convey what decisions are prominent given certain results.
Our visualizer uses trellis plots and supplements it with brushing to show how decisions contribute to particular results.

Finally, prior work relies on various strategies to infer whether a hypothesized effect occurs given a multiverse. 
The simplest approach is counting the fraction of universes having a significant p-value~\cite{steegen2016, cesario2019} and/or an effect with the same sign~\cite{dejonckheere2018}. 
Young and Holsteen~\cite{young2017} calculate a robustness ratio analogous to the \textit{t}-statistic.
Simonsohn et al.~\cite{simonsohn2015} compare the actual multiverse results to a null distribution obtained from randomly shuffling the variable of interest.
We build upon Simonsohn's approach and use weighted model averaging based on model fit quality~\cite{yao2018} to aggregate uncertainty across universes.

\vv{While multiverse analysis is a recent concept, prior work has developed visual analytics approaches for similar problems.
For example, multiverse analysis fits into the broader definition of parameter space analysis~\cite{booshehrian2012, sedlmair2014}, a concept originally proposed for understanding inputs and outputs of simulation models.
Visual analytics systems for preprocessing time-series data~\cite{bernard2012, bernard2019} also propose ways to generate and visualize alternative results, for example via superposition.}



\subsection{Authoring Alternative Programs and Designs}
Analysts often manage alternatives from exploratory work by duplicating code snippets and files, but these ad-hoc variants can be messy and difficult to keep track of~\cite{kery2017, guo2012thesis}.
Provenance tracking tools, especially those with enhanced history interactions~\cite{kery2017, kery2018}, provide a mechanism to track and restore alternative versions.
In Variolite~\cite{kery2017}, users select a chunk of code directly in an editor to create and version alternatives.
We also allow users to insert local alternatives in code, but instead of assuming that users interact with one version at a time, we generate multiple variants mapping to possible combinations of alternatives.

A related line of work supports manipulating multiple alternatives simultaneously.
Techniques like subjunctive interfaces~\cite{lunzer1998, lunzer1999} and Parallel Pies~\cite{terry2004} embed and visualize multiple design variants in the same space, and Parallel Pies allows users to edit multiple variants in parallel.
Juxtapose~\cite{hartmann2008} extends the mechanism to software development, enabling users to author program alternatives as separate files and edit code duplicates simultaneously with linked editing.
A visualization authoring tool for responsive design~\cite{hoffswell2020} also enables simultaneous editing across variants. 
Our DSL uses a centralized template such that edits in the shared portion of code affect all variants simultaneously. 
\section{Design Requirements}

Our overarching goal is to make it easier for researchers to conduct multiverse analyses.
From prior literature and our past experiences, we identify barriers in authoring a multiverse and visualizing its results, and subsequently identify tasks that our tools should support. 

\subsection{Requirements for Authoring Tool}
\label{task:dsl}

As noted in prior work~\cite{dragicevic2019, liu2020-paths}, specifying a multiverse is tedious.
This is primarily because a multiverse is composed of many forking paths, yet non-linear program structures are not well supported in conventional tools~\cite{rule2018}.
One could use a separate script per analytic path, such that it is easy to reason with an individual variant, but 
these variants are redundant and difficult to maintain~\cite{kery2017}. 
Alternatively, one could rely on control flows in a single script to simulate the nonlinear execution, but it is hard to selectively inspect and rerun a single path, and deeply nested control flows are thought to be a software development anti-pattern~\cite{mcconnell2004}.
Instead, a tool should eliminate the need to write redundant code and custom control flows, while allowing analysts to simultaneously update variants and reason with a single variant.
Compared to arbitrary non-linear paths from an iterative exploratory analysis, the forking paths in multiverses are usually highly systematic.  
We take advantage of this characteristic, and account for other scenarios common in existing multiverse analyses.
We distill the following design requirements:

R1: \textbf{Multiplexing}. Users should be able to specify a multiverse by writing the shared portion of the analysis source code along with analytic decisions, while the tool creates the forking paths for them. 
Users should also be able to reason about a single universe and update all universes simultaneously.

R2: \textbf{Decision Complexity}. Decisions come in varying degrees of complexity, from simple value replacements (\eg cutoffs for excluding outliers) to elaborate logic requiring multiple lines of code to implement. 
The tool should allow succinct ways to express simple value replacements while at the same time support more complex decisions.

R3: \textbf{Procedural Dependency}. Existing multiverses~\cite{steegen2016, crede2017} contain \emph{procedural dependencies}~\cite{liu2020-paths}, in which a downstream decision only exists if a particular upstream choice is made. 
For example, researchers do not need to choose priors if using a Frequentist model instead of a Bayesian model.
The tool should support procedural dependencies.

R4: \textbf{Linked Decisions}. Due to idiosyncrasies in implementation, the same conceptual decision can manifest in multiple forms.
For example, the same set of parameters can appear in different formats to comply with different function APIs.
Users should be able to specify different implementations of a high-level decision.

R5: \textbf{Language Agnostic}. Users should be able to author their analysis in any programming languages, as potential users are from various disciplines adopting different workflows and programming languages.

\figureWorkflow
\subsection{Requirements for Visual Analysis System}
\label{task:visualizer}

After executing all analytic paths in a multiverse to obtain corresponding results, researchers face challenges interpreting the results collectively.
The primary task in prior work (\autoref{sec:related-work}) is understanding the robustness of results across all reasonable specifications.
If robustness checks indicate conflicting conclusions, a natural follow-up task is to identify what decisions are critical to reaching a particular conclusion or what decisions produce large variations in results.

We also propose new tasks to cover potential blind spots in prior work.
First, besides point estimates, a tool should convey appropriate uncertainty information to help users gauge the end-to-end uncertainty caused by both sampling and decision variations, and compare the variance between conditions.
Second, it is important to assess the model fit quality to distinguish trustworthy models from the ones producing questionable estimates.
Uncertainty information and fit issues become particularly important during statistical inference. Users should be able to propagate uncertainty in the multiverse to support judgments about the overall reliability of effects, and they should be able to refine the multiverse to exclude models with fit issues before making inferences.

We identify six tasks that our visual analysis system should support: 

\begin{itemize}
	\item T1: \textit{Decision Overview} -- gain an overview of the decision space to understand the multiverse and contextualize subsequent tasks.
	\item T2: \textit{Robustness Overview} -- gauge the overall robustness of findings obtained through all reasonable specifications.
	\item T3: \textit{Decision Impacts} -- identify what combinations of decisions lead to large variations in outcomes, and what combinations of decisions are critical in obtaining specific outcomes. 
	\item T4: \textit{Uncertainty} -- assess the end-to-end uncertainty as well as uncertainty associated with individual universes.
	\item T5: \textit{Model Fit} -- assess the model fit quality of individual universes to distinguish trustworthy models from questionable ones.
	\item T6: \textit{Inference} -- perform statistical inference to judge how reliable the hypothesized effect is, while accounting for model quality.
\end{itemize}

\vv{Besides the tasks, our system should also support the following data characteristics (S1) and types of statistical analyses (S2).
First, our visual encoding should be scalable to large multiverses and large input datasets.
This is because the multiverse size increases exponentially with the number of decisions, with the median size in practice being in the thousands~\cite{liu2020-paths}.
The input datasets might also have arbitrarily many observations.
Second, we should support common simple statistical tests in HCI research~\cite{phelan2019}, including ANOVA and linear regressions.
}

\subsection{Workflow}

\vv{We propose a general workflow for multiverse analysis with four stages (\autoref{fig:workflow}).
In this workflow, users \textit{author} the multiverse specification, \textit{explore} the results, \textit{refine} the multiverse by pruning universes with poor model quality, and make \textit{inference}.
Users should be free to cycle between the first three stages, because upon exploring the results, users might discover previously overlooked alternatives, or notice that certain decisions are poorly suited for inference.
In this case, they might iterate on their multiverse specification to include only decisions resulting in universes that seem ``reasonable''.
However, once users proceed to the \textit{inference} stage, they should not return to any of the prior stages.}

\section{The Boba DSL}
\label{sec:dsl}

We design a domain-specific language (DSL) to aid the authoring of multiverse analyses.
\vv{The DSL formally models an analysis decision space, providing critical structure that the visual analysis system later leverages.}
With the DSL, users annotate the source code of their analysis to indicate decision points and alternatives, and provide additional information for  procedural dependencies between decisions.
The specification is then compiled to a set of universe scripts, each containing the code to execute one analytic path in the multiverse.
An example Boba specification for a small multiverse is shown in \autoref{fig:syntax}.

\subsection{Language Constructs}

The basic language primitives in the Boba DSL consist of source code, placeholder variables, blocks, constraints, and code graphs.

\figureSyntax

\fakeTitle{Source Code}
The most basic ingredient of an annotated script is the source code (\autoref{fig:syntax}a, black text).
The compiler treats the source code as a string of text, which according to further language rules will be synthesized into text in the output files.
As the compiler is agnostic about the semantics of the source code, users are free to write the source code in any programming language~(R5).

\fakeTitle{Placeholder Variables}
Placeholder variables are useful to specify decisions points consisting of simple value substitution (R2).
To define a placeholder variable, users provide an identifier and a set of possible alternative values that the variable can take up (\autoref{fig:syntax}a, blue text).
To use the variable, users insert the identifier into any position in the source code.
During synthesis, the compiler removes the identifier and replaces it with one of its alternative values.
Variable definition may occur at the same place as its usage (\autoref{fig:syntax}a) or ahead of time inside the config block (supplemental Fig. 2).

\fakeTitle{Code Blocks}
Code blocks (\autoref{fig:syntax}a, pink text) divide the source code into multiple snippets of one or more lines of code, akin to cells in a computational notebook.
A block can be a \textit{normal block} (\autoref{fig:syntax}a, block A), or a \textit{decision block} (\autoref{fig:syntax}a, block M) with multiple versions.
The content of a normal block will be shared by all universes, whereas only one version of the decision block will appear in a universe.
Decision blocks allow users to specify alternatives that require more elaborate logic to define (R2).
In the remainder of \autoref{sec:dsl}, \textit{decision points} refer to placeholder variables and decision blocks.

With the constructs introduced so far, a natural way to express procedural dependency (R3) is to insert a placeholder variable in some, but not all versions of a decision block.
For example, in \autoref{fig:syntax}, the variable \texttt{brm\_family} only exists when \texttt{bayesian} of block M is chosen.


\fakeTitle{Constraints}
By default, Boba assumes all combinations between decision points are valid.
Constraints allow users to express dependencies between decision points, for example infeasible combinations, which will restrict the universes to a smaller set.
Boba supports two types of constraints: procedural dependencies (R3) and linked decisions (R4).

A procedural dependency constraint is attached to a decision point or one of its alternatives, and has a conditional expression to determine when the decision/alternative should exist (\autoref{fig:dsl-example}b, orange text).
Variables within the scope of the conditional expression are declared decision points, and the values are the alternatives that the decision points have taken up.
For example, the first constraint in \autoref{fig:dsl-example}b indicates that \texttt{ECL computed} is not compatible with \texttt{NMO reported}.

The second type of constraint allows users to \textit{link} multiple decision points,
indicating that these decision points are different manifestations of a single conceptual decision (R4, see supplemental Fig. 2).
Linked decisions have one-to-one mappings between their alternatives, such that the $i$-th alternatives are chosen together in the same universe.
One-to-one mappings can also be expressed using multiple procedural dependencies, but linked decisions make them easier to specify.

\fakeTitle{Code Graph}
Users may further specify the execution order between code blocks as a directed acyclic graph (DAG), where a parent block executes before its child.
To create a universe, the compiler selects a linear path from the start to the end, and concatenates the source code of blocks along the path.
Branches in the graph represent alternative paths that appear in different universes.
Users can flexibly express complex dependencies between blocks with the graph, including procedural dependencies (R3).
For example, to indicate that block \texttt{prior} should only appear after block \texttt{bayesian} but not block \texttt{frequentist}, the user simply makes \texttt{prior} a descendant of \texttt{bayesian} but not \texttt{frequentist}.

\subsection{Compilation and Runtime}

The compiler parses the input script, computes compatible combinations between decisions, and generates output scripts.
More details about compilation are in the supplemental material.
Besides executable universe scripts, the compiler also outputs a summary table that keeps track of all the decisions made in each universe, along with other intermediate data that can be ingested into the Boba Visualizer.

Boba infers the language of the input script based on its file extension and uses the same extension for output scripts.
These output scripts might be run with the corresponding script language interpreter.
Universe scripts log the results into separate files, which will be merged together after all scripts finish execution.
\vv{Each universe must output a point estimate, along with other optional data such as a p-value, a model quality metric, or a set of sampled estimates to represent uncertainty.}
As the universe scripts are responsible for computations such as extracting point estimates and computing uncertainty, we provide language-specific utilities for a common set of model types to generate these visualizer-friendly outputs.
We also provide a command-line tool for users to (1) invoke the compiler, (2) execute the generated universe scripts, (3) merge the universe outputs, and (4) invoke the visualizer as a local server reading the intermediate output files.

\figureDslExample
\subsection{Example: Replicating a Real-World Multiverse}
We use a real-world multiverse example~\cite{steegen2016} to illustrate how the Boba DSL eliminates the need for custom control flows otherwise required for authoring a multiverse in a single script.
The multiverse, originally proposed by Steegen~\etal~\cite{steegen2016}, contains five decisions and a procedural dependency.
\autoref{fig:dsl-example}a shows a markup of the R code implemented by the original authors (we modified the lines in purple to avoid computing infeasible paths).
The script starts with five nested for-loops (yellow highlight) to repeat the analysis for every possible combination of the five decisions.
Then, depending on the indices of the current decisions, the authors either index into an array, or use if-statements to define alternative program behaviors (blue highlight).
Finally, to implement a procedural dependency, it is necessary to skip the current iteration when incompatible combinations occur (purple highlight).

The resulting script has multiple issues.
\vv{First, the useful snippets defining multiverse behavior start at five levels of nesting at minimum.
Such deeply nested code is often considered to be hard to read~\cite{mcconnell2004}.
Second, nested control flows are not easily amenable to parallel execution.
Third, without additional error-handling mechanisms, an error in the middle will terminate the program before any results are saved.}

The corresponding specification in the Boba DSL is shown in \autoref{fig:dsl-example}b.
The script starts directly with the preprossessing code shared by all universes.
It then uses decision code blocks to define alternative snippets in decision \texttt{NMO} and \texttt{ECL}, and uses a placeholder variable to simulate the value array for a simpler decision \texttt{F}.
It additionally specifies constraints (orange text) to signal incompatible paths.
Compared to \autoref{fig:dsl-example}a, this script reduces the amount of boilerplate code needed for control-flows \vv{and does not require any level of nesting.} 
The script compiles to 120 separate files. 
Errors in one universe no longer affect the completion of others due to distributed execution, \vv{it is trivial to execute universes in parallel}, and users can selectively review and debug a single analysis.

\section{The Boba Visualizer}

Next, we introduce Boba Visualizer, a visual analysis system for interpreting the outputs from all analysis paths.
We present the system features and design choices in a fictional usage scenario where Emma, an HCI researcher, uses the visualizer to explore a multiverse on data collected in her experiment.
We construct the multiverse based on how the authors of a published research article~\cite{li2019-reader-view} might analyze their data, but the name ``Emma'' and her workflow are fictional.

\figureSystemStartup

\vv{Emma runs an experiment to understand whether ``Reader View'' -- a modified web page layout -- improves reading speed for individuals with dyslexia.
She assigns participants 
to use 
Reader View or standard websites, measures their reading speed, and collects other variables such as accuracy, device, and demographic information.
She plans to build a model with reading speed as the dependent variable.
To check whether her conclusion depends on idiosyncratic specifications, Emma identifies six analytic decisions, 
including the \texttt{device} type and \texttt{accuracy} cutoff used to filter participants, ways to operationalize \texttt{dyslexia}, the statistical \texttt{model}, and its \texttt{random} and \texttt{fixed} terms.}
She then writes a multiverse specification in the Boba DSL, compiles it to 216 analysis scripts, 
and runs all scripts to obtain a set of effect sizes. 
She loads these outputs into the Boba Visualizer.

\subsection{Outcome View}
\q{On system start-up, Emma sees an overview distribution of point estimates from all analyses (\autoref{fig:system-startup}b). The majority of the coefficients are positive, but a smaller peak around zero suggests no effect.}

The outcome view visualizes the final results of the multiverse, including point estimates (\eg model coefficient of \texttt{reader view}, the independent variable encoding experimental conditions) and uncertainty information.
By default, the chart contains outcomes from all universes in order to show the overall robustness of the conclusion (T2).

Boba visualizes one point estimate from each universe using a density dot plot~\cite{wilkinson1999} (\autoref{fig:system-startup}b, blue dots).
The x-axis encodes the magnitude of the estimate; dots in the same bin are 
stacked along the y-axis.
To accommodate large multiverses (S1), we allow dots to overlap along the y-axis, which 
represents count.
Density dot plots more accurately depict gaps and outliers in data than histograms~\cite{wilkinson1999}.
One-to-one mapping between dots and universes affords direct manipulation interactions such as brushing and details-on-demand, as we 
introduce later. 

Boba visualizes end-to-end uncertainty from both sampling and decision variations (T4) as a background area chart (\autoref{fig:system-startup}b, gray area).
When the uncertainty introduced by sampling variations is negligible, the area chart follows the dot plot distribution closely.
In contrast, the mismatch of the two distributions in \autoref{fig:system-startup}b indicates considerable sampling uncertainty.
We compute the end-to-end uncertainty by aggregating over modeling uncertainty from all universes.
Specifically, we first calculate $\hat{f}(x) = \sum_{i=1}^{N}f_i(x)$, where $f_i(x)$ is the sampling distribution of the $i$-th universe, and $N$ is the overall multiverse size.
Then, we scale the height of the area chart such that the total area under $\hat{f}(x)$ is approximately the same as the total area of dots in the dot plot.

Besides aggregated uncertainty, Boba allows users to examine uncertainty from individual universes (\autoref{fig:system-curves}).
In a dropdown menu, users can switch to view the probability density functions (PDFs) or cumulative distribution functions (CDFs) of all universes.
\vv{A PDF is a function that maps the value of a random variable to its likelihood, whereas a CDF gives the area under the PDF.}
In both views, we draw a cubic basis spline for the PDF or CDF per universe, and reduce the opacity of the curves to visually ``merge'' the curves within the same space.
There is again a one-to-one mapping between a visual element and a universe to afford interactions.
To help connect point estimates and uncertainty, we draw a strip plot of point estimates beneath each PDFs/CDFs chart (\autoref{fig:system-curves}, blue dashes), and show the corresponding sampling distribution PDF when users mouse over a universe in the dot plot.

\figureFacetBrush
\figureCurves
\subsection{Decision View}
\q{As the overall outcome distribution suggests conflicting conclusions, Emma wants to investigate what decisions lead to changes in results. She first familiarizes herself with the available decisions.}

The decision view shows a graph of 
analytic decisions 
in the multiverse, along with their order and dependencies (\autoref{fig:system-startup}a),
helping users understand the decision space and inviting further exploration (T1). 

We adapt the design of Analytic Decision Graphs~\cite{liu2020-paths} to show decisions in the context of the analysis process.
Nodes represent decisions and edges represent the relationships between decisions: light gray edges encode \textit{temporal order} (the order that decisions appear in analysis scripts) and black edges encode \textit{procedural dependencies}.
To avoid visual clutter, we aggregate the information about alternatives, using the size of a node to represent the number of alternatives and listing a few example alternative values besides a node.
Compared to viewing decisions in isolation, this design additionally conveys the analysis pipeline to help users better reason with the ramifications of a decision.

The underlying data structure for the graph is inferred from the Boba DSL specification. 
We infer decision names from variable identifiers.
We extract temporal order as the order that decision points are first used in the specification, and detect procedural dependencies from user-specified constraints and code graph structure.
After we extract the data structure, we apply a Sugiyama-style~\cite{sugiyama1981} flow layout algorithm, as implemented in Dagre~\cite{dagre}, to compute the graph layout. 

\subsubsection{Sensitivity}
\q{When viewing the decision graph, Emma notes a sensitive decision ``Device''  which is highlighted in a darker color (\autoref{fig:system-startup}a).}

To highlight decisions that lead to large changes in analysis outcomes (T3), we compute the marginal sensitivity per decision and color the nodes using a sequential color scale. The color encoding helps draw the user's attention to consequential decisions to aid initial exploration.

Boba implements two methods for estimating sensitivity.
The first method is based on the F-Test in one-way ANOVA, which quantifies how much a decision shifts the means of results compared to variance (details in supplemental material). 
The second method uses the Kolmogorov--Smirnov (K--S) statistic, a non-parametric method to quantify the difference between two distributions.
We first compute pairwise K--S statistics between all pairs of alternatives in decision $D$:

\begingroup
\setlength\abovedisplayskip{3pt}
\setlength\belowdisplayskip{3pt}
\[K = \left\{ \sup_x|f_i(x) - f_j(x)|: i, j \in {S \choose 2} \right\}\]
\endgroup 
where $f_i(x)$ is the empirical distribution function of results following the $i$-th alternative, and $S = \{1, 2, ..., k \}$ where $k$ is the number of alternatives in $D$.
We then take the median of $K$ as the sensitivity score $s_D$. In both methods, we map $s_D$ to a single-hue color ramp of blue shades.
As the F-Test relies on variance, which is not a reasonable measure for dispersion of some distributions, Boba uses the nonparametric K--S statistic by default.
Users can override the default in the config file.


\subsection{Facet and Brushing}
\q{Seeing that the decision ``Device'' has a large impact, Emma clicks on the node to further examine how results vary (\autoref{fig:system-facet}a). She finds that one condition exclusively produces point estimates around zero (\autoref{fig:system-facet}b) and it also has a much larger variance (\autoref{fig:system-curves}).  }

Clicking a node in the decision graph facets the outcome distribution into a trellis plot, grouping subsets of universes by shared decision alternatives. 
This allows users to systematically examine the trends and patterns caused by a decision (T3).
Akin to the overall outcome distribution, users can toggle between point estimates and uncertainty views to compare the variance between conditions.
The trellis plot can be further divided on a second decision by shift-clicking a second node to show the interaction between two decisions.
With faceting, users may comprehensively explore the data by viewing all univariate and bivariate plots. 
Sensitive decisions are automatically highlighted, so users might quickly find and examine consequential decisions as well. 

\breakline
\q{What decisions lead to negative estimates? Emma brushes negative estimates in a subplot (\autoref{fig:system-facet}c) and inspects option ratios (\autoref{fig:system-facet}d).}

Brushing a region in the outcome view updates the option ratio view. 
The option ratio view shows percentages of decision options to reveal dominating alternatives that produce specific results (T3). 

The option ratio 
view visualizes each decision as a stacked bar chart, \vv{
where bar segment length 
encodes the percentage of results coming from an alternative.}
When the user brushes a range of results, the bars are updated accordingly to reflect changes, and dominating alternatives (those having a higher percentage than default) are highlighted.
For example, Emma notices that the \texttt{lmer} model (\ie linear mixed-effect model in R) and two sets of fixed effects are particularly responsible for the negative outcomes in \autoref{fig:system-facet}c.
We color the bar segments using a categorical color scale to make bars visually distinguishable.
\vv{Decisions used to divide a trellis plot are marked by a striped texture, as each trellis subplot only contains one alternative by definition.}


\subsection{Model Fit View}\label{sec:model-fit}
\q{Now that Emma understands what decisions lead to null effects, she wonders if these results are from trustworthy models. She changes the color-by field to get an overview of model fit quality (\autoref{fig:system-prune}a) and sees that the universes around zero have a poorer fit. She then uses a slider to remove universes that fail to meet a quality threshold (\autoref{fig:system-prune}b).}

Boba enables an overview of model fit quality across all universes (T5) by coloring the outcome view with a model quality metric (\autoref{fig:system-prune}a).
We use normalized root mean squared error (NRMSE) to measure model quality and map NRMSE to a single-hue colormap of blue shades where a darker blue indicates a better fit.

To obtain NRMSE, we first compute the overall mean squared prediction error (MSE) from a $k$-fold cross validation:
\begingroup
\setlength\abovedisplayskip{3pt}
\setlength\belowdisplayskip{3pt}
\[MSE = \dfrac{1}{k}\sum_{j=1}^k \dfrac{1}{n_j} \sum_{i=1}^{n_j}(y_i - \hat{y_i})^2 \]
\endgroup 
where $k$ is the number of folds (we set $k=5$ in all examples), $n_j$ is the size of the test set in the $j$-th fold, $y_i$ is the observed value, and $\hat{y_i}$ is the predicted value.
We then normalize the MSE by the span of the maximum $y_{max}$ and minimum $y_{min}$ values of the observed variable:
\begingroup
\setlength\abovedisplayskip{3pt}
\setlength\belowdisplayskip{3pt}
\[NRMSE = \sqrt{MSE} / (y_{max} - y_{min})\]
\endgroup 

We use $k$-fold cross validation~\cite{Vehtari2017} because metrics such as Akaike Information Criterion cannot be used to compare model fit across classes of models (e.g., hierarchical vs. linear)~\cite{Gelman2014-information_criteria}.
Prior work shows that cross validation performs better in estimating predictive density for a new dataset than information criteria~\cite{Vehtari2017}, suggesting that it is a better approximation of out-of-sample predictive validity.

\figurePrune
\figureInference

\breakline
\q{To further investigate model quality, Emma drills down to individual universes by clicking a dot in the outcome view. She sees in the model fit view (\autoref{fig:teaser}e) that a model gives largely mismatched predictions.}

Clicking a result in the outcome view populates the model fit view with visual predictive checks, which show how well predictions from a given model replicate the empirical distribution of observed data~\cite{Gelman2003}, allowing users to further assess model quality (T5).
The model fit visualization juxtaposes violin plots of the observed data and model predictions to facilitate comparison of the two distributions (see \autoref{fig:teaser}e). 
Within the violin plots, we overlay observed and predicted data points as centered density dot plots to help reveal discrepancies in approximation due to kernel density estimation. 
When the number of observations is large (S1), we plot a representative subset of data, sampled at evenly spaced percentiles, as centered quantile dotplots~\cite{kay2016}.
As clicking individual universes can be tedious, the model fit view suggests additional universes that have similar point estimates to the selected universe.

\figureMortgage

\subsection{Inference}

\q{After an in-depth exploration, Emma proceeds to the final step, asking ``given the multiverse, how reliable is the effect?'' She confirms a warning dialog to arrive at the inference view (\autoref{fig:system-inference}).}

To support users in making inference and judging how reliable the hypothesized effect is (T6), Boba provides an inference view at the end of the analysis workflow, after users have engaged in exploration.
Once in the inference view, all earlier views and interactions are inaccessible to avoid multiple comparison problems~\cite{zgraggen2018} arising from repeated inference.
The inference view contains different plots depending on the outputs from the authoring step, so that users can choose between robust yet computationally-expensive methods and simpler ones.

A more robust inference utilizes the null distribution -- the expected distribution of outcomes when the null hypothesis of no effect is true.
In this case, the inference view shows an aggregate plot followed by a detailed plot (\autoref{fig:system-inference}ab).
The aggregate plot (\autoref{fig:system-inference}a) compares the null distribution (red) to possible outcomes of the actual multiverse (blue) across sampling and decision variations.
The detailed plot (\autoref{fig:system-inference}b) shows point estimates (colored dots) against 95\% confidence intervals representing null distributions (gray lines) for each universe. 
Each point estimate is orange if it is outside the range, or blue otherwise.
Underneath both plots, we provide
descriptions (supplemental Fig. 1) to guide users in interpretation:
For the aggregate plot, we prompt users to compare the distance between the averages of the two densities to the spread.
For the detailed plot, we count the number of universes with the point estimate outside its corresponding range.
If the null distribution is unavailable, Boba shows a simpler aggregate plot (\autoref{fig:system-inference}c) where the expected effect size under the null hypothesis is marked with a red line.

To compute the null distribution, 
we permute the data with random assignment~\cite{simonsohn2015}. 
Specifically, we shuffle the column with the 
independent
variable (\texttt{reader view} in this case) $N$ times, run the multiverse of size $M$ on each of the shuffled datasets, and obtain $N \times M$ point estimates.
As 
\texttt{reader view} and \texttt{speed}
are independent 
in the shuffled datasets, 
these $N \times M$ point estimates constitute the null distribution.

In addition, Boba enables users to propagate concerns in model fit quality to the inference view in two possible ways.
The first way employs a model averaging technique called \textit{stacking}~\cite{yao2018} to take a weighted combination of the universes according to their model fit quality.
The technique learns a simplex of weights, one for each universe model, via optimization that maximizes the log-posterior-density of the held-out data points in a $k$-fold cross validation.
Boba then takes a weighted combination of the universe distributions to create the aggregate plot.
While stacking provides a principled way to approach model quality, it can be computationally expensive.
As an alternative, Boba excludes the universes below the model quality cutoff users provide in~\autoref{sec:model-fit}.
The decisions of the cutoff and whether to omit the universes are made before a user enters the inference view.


\section{Case Studies}

We evaluate Boba through a pair of analysis case studies, where we implement the multiverse using the Boba DSL and interpret the results using the Boba Visualizer.
The supplemental material contains the Boba specifications of both examples, additional figures, and a video demonstrating all the interactions described below.

\figureHurricane
\subsection{Case Study: Mortgage Analysis}
The first case study demonstrates how analysts might quickly arrive at insights provided by summary statistics in prior work, while at the same time gaining a richer understanding of robustness patterns.
We also show that by incorporating uncertainty and model fit checks, Boba surfaces potential issues that prior work might have neglected.

Young \etal~\cite{young2017} propose a multimodel analysis approach to gauge whether model estimates are robust to alternative model specifications.
Akin to the philosophy of multiverse analysis, the approach takes all combinations of possible control variables in a statistical model.
The outputs are multiple summary statistics,  including (1) an overall \textit{robustness ratio}, (2) \textit{uncertainty} measures for sampling and modeling variations, and (3) metrics reflecting the \textit{sensitivity} of each variable.

\vv{As an example, the authors present a case study on mortgage lending, asking ``are female applicants more likely to be approved for a mortgage?''
They use a dataset of publicly disclosed loan-level information about mortgage, and fit a linear regression model with mortgage application acceptance rate as the dependent variable and female as one independent variable.
In their multimodel analysis, they test different control variables capturing demographic and financial information.}
The resulting summary statistics indicate that the estimate is not robust to modeling decisions with large end-to-end uncertainty, and two control variables, \textit{married} and \textit{black}, are highly influential.
These summary statistics provide a powerful synopsis, but may fail to convey more nuanced patterns in result distributions.
The authors manually create additional visualizations to convey interesting trends in data, for instance the estimates follow a multimodal distribution.
These visualizations, though necessary to provide a richer understanding of model robustness, are ad-hoc and not included in the software package.


We replicate the analysis in Boba.
The Boba DSL specification simply consists of eight placeholder variables, each indicating whether to include a control variable in the model formula.
Then, we compile the specification to 256 scripts, run them all, and start the Boba Visualizer.

We first demonstrate that the default views in the Boba Visualizer afford similar insights on uncertainty, robustness, and decision sensitivity.
Upon launching the visualizer, we see a decision graph and an overall outcome distribution (\autoref{fig:mortgage}).
The decision view (\autoref{fig:mortgage}a) highlights two sensitive decisions, \textit{black} and \textit{married}.
The outcome view (\autoref{fig:mortgage}b) shows that the point estimates are highly varied with conflicting implications.
The aggregated uncertainty in the outcome view (\autoref{fig:mortgage}b, background gray area) has a wide spread, suggesting that the possible outcomes are even more varied when taking both sampling and decision variability into account.
These observations agree with the summary metrics in previous work, though Boba uses a different, non-parametric method to quantify decision sensitivity, as well as a different method to aggregate end-to-end uncertainty.

The patterns revealed by ad-hoc visualizations in previous work are also readily available in the Boba Visualizer, either in the default views or with two clicks guided by prominent visual cues.
The default outcome view (\autoref{fig:mortgage}b) shows that the point estimates follow a multimodal distribution with three separate peaks.
Clicking the two highlighted (most sensitive) nodes in the decision view (\autoref{fig:mortgage}a) produces a trellis plot (\autoref{fig:mortgage}c), where each subplot contains only one cluster.
From the trellis plot, it is evident that the leftmost and rightmost peaks in the overall distribution come from two particular combinations of the influential variables.
Alternatively, users might arrive at similar insights by brushing individual clusters in the default outcome view.

Finally, the uncertainty and model fit visualizations in Boba surface potential issues that previous work might have overlooked.
First, though the point estimates in \autoref{fig:mortgage}b fall into three distinct clusters, the aggregated uncertainty distribution appears unimodal despite a wider spread.
The PDF plot (\autoref{fig:mortgage}e) shows that sampling distribution from one analysis typically spans the range of multiple peaks, thus explaining why the aggregated uncertainty is unimodal.
These observations suggest that the multimodal patterns exhibited by point estimates are not robust when we take sampling variations into account.
Second, we assess model fit quality by clicking a dot in the outcome view and examining the model fit view (\autoref{fig:mortgage}d).
As shown in \autoref{fig:mortgage}d, while the observed data only takes two possible values, the linear regression model produces a continuous range of predictions. 
It is clear from this visual check that an alternative model, for example logistic regression, is more appropriate than the original linear regression models, and we should probably interpret the results with skepticism given the model fit issues.
These observations support our arguments in \autoref{task:visualizer} that uncertainty and model fit are potential blind spots in prior literature.

\subsection{Case Study: Female Hurricanes Caused More Deaths?}

Next, we replicate another multiverse where Simonsohn~\etal~\cite{simonsohn2015} challenged a previous study~\cite{jung2014}.
\vv{The original study~\cite{jung2014} reports that hurricanes with female names have caused more deaths, presumably because female names are perceived as less threatening and lead to less preparation.
The study used archival data on hurricane fatalities and regressed death count on femininity.}
However, the study led to a heated debate on proper ways to conduct the data analysis.
To understand if the conclusion is robust to alternative specifications, Simonsohn~\etal identified seven analytic decisions, including alternative ways to exclude outliers, operationalize femininity, select the model type, and choose covariates. 
They then conducted a multiverse analysis and interpreted the results in a visualization called a \textit{specification curve}.

We build the same multiverse using these seven analytic decisions in Boba.
In the Boba DSL specification, we use a decision block to specify two alternative model types: negative binomial regression versus linear regression with log-transformed deaths as the dependent variable.
The rest of the analytic decisions are placeholder variables that can be expressed as straightforward value substitutions.
However, the two different model types lead to further differences in extracting model estimates.
For example, we must invert the log-transformation in the linear model to obtain predictions in the original units.
We create additional placeholder variables for implementation differences related to model types and link them with the model decision block.
The specification compiles to 1,728 individual scripts.

We then interpret the results using the Boba Visualizer.
As shown in the overview distribution (\autoref{fig:hurricane}a), the majority of point estimates support a small, positive effect (female hurricanes lead to more deaths, and the extra deaths are less than 20), while some estimates suggest a larger effect.
A small fraction of results have the opposite sign.

What analytic decisions are responsible for the variations in the estimates?
The decision view indicates that multiple analytic decisions might be influential (\autoref{fig:hurricane}a).
We click on the relatively sensitive decisions, \textit{outliers}, \textit{damage} and \textit{model}, to examine their impact.
In the corresponding univariate trellis plots (supplemental Fig. 3), certain choices tend to produce larger estimates, such as not excluding any outliers, using raw damage instead of log damage, and using negative binomial regression.
However, in each of these conditions, a considerable number of universes still support a smaller effect, suggesting that it is not a single analytic decision that leads to large estimates.

Next, we click on two influential decisions to examine their interaction.
In the trellis plot of \textit{model} and \textit{damage} (\autoref{fig:hurricane}b), one combination (choosing both log damage and negative binomial model) produces mostly varied estimates without a dominating peak next to zero.
Brushing the large estimates in another combination (raw damage and linear model) indicates that these results are coming from specifications that additionally exclude no outliers.
Removing these decision combinations will eliminate the possibility of obtaining a large effect.

But do we have evidence that certain outcomes are less trustworthy?
We toggle the color-by drop-down menu so that each universe is colored by its model quality metric (\autoref{fig:hurricane}b).
The large estimates are almost exclusively coming from models with a poor fit.
We further verify the model fit quality by picking example universes and examining the model fit view (\autoref{fig:hurricane}c).
The visual predictive checks confirm issues in model fit, for example the models fail to generate predictions smaller than 3 deaths, while the observed data contains plenty such cases.

Now that we have reasons to be skeptical of the large estimates, the remaining universes still support a small, positive effect.
How reliable is the effect?
We proceed to the inference view to compare the possible outcomes in the observed multiverse and the expected distribution under the null hypothesis (\autoref{fig:hurricane}d).
The two distributions are different in terms of mode and shape, yet they are highly overlapping, which suggests the effect is not reliable.
The detail plot depicting individual universes (supplemental Fig. 1) further confirms this observation. Out of the entire multiverse, only 3 universes have point estimates outside the 2.5th and 97.5th percentile of the corresponding null distribution.

\section{Discussion}

Through the process of designing, building, and using Boba, we gain insights into challenges that multiverse analysis poses for software designers and users. We now reflect on these challenges and additional design opportunities for supporting multiverse analysis.

While Boba is intended to reduce the gulf of execution for multiverse analysis, conducting a multiverse analysis still requires statistical expertise.
\vv{The target users of our current work are experienced researchers and statisticians.} 
Future work might attempt to represent expert statistical knowledge to lower the barriers for less experienced users.
One strategy is to represent analysis goals in higher-level abstractions, from which appropriate analysis methods might be synthesized~\cite{jun2019}.
Another is to guide less experienced users through key decision points and possible alternatives~\cite{liu2020-paths}, starting from an initial script.

Running all scripts produced by Boba can be computationally expensive due to their sheer number.
Boba already leverages parallelism, executing universes across multiple processes.
Still, scripts often perform redundant computation and the compiler may produce prohibitively many scripts. 
Future work should include optimizing multiverse execution, for example caching shared computation across universes, or efficiently exploring decision spaces via adaptive sampling.

As a new programming tool, Boba requires additional support to increase its usability, including code editor plugins, debugging tools, documentation, and community help.
\vv{In this paper we assess the feasibility of Boba, with the understanding that its usability will need to be subsequently evaluated.} 
Currently, Boba specifications are compiled into scripts in a specific programming language, so users can leverage existing debugging tools for the corresponding language. 

However, debugging analysis scripts becomes difficult at the scale of a multiverse because a change that fixes a bug in one script might not fix bugs in others.
When we attempt to run a multiverse of Bayesian regression models, for example, models in multiple universes do not converge for a variety of reasons including problems with identifiability and difficulties sampling parameter spaces with complex geometries.
These issues are common in Bayesian modeling workflows and must be resolved by adjusting settings, changing priors, or reparameterizing models entirely.
At the scale of multiverse analysis, debugging this kind of model fit issue is particularly difficult because existing tools for diagnostics and model checks (e.g., trace and pairs plots) are designed to assess one model at a time.
While this points to a need for better debugging and model diagnostic tools in general, it also suggests that these tools must be built with a multiplexing workflow in mind if they are going to facilitate multiverse analysis.

Analysts must take care when reviewing and summarizing multiverse results, as a multiverse is not a set of randomly drawn, independent specifications.
In general, the Boba Visualizer avoids techniques that assume universe results are independent and identically distributed.
A possible venue for future work is to explicitly account for statistical dependence among universes to remove potential bias.
Boba might also do more to aid the communication of results, for example helping to produce reports that communicate multiverse results~\cite{dragicevic2019}.

Previous approaches to multiverse analysis have largely overlooked the quality of model fit, focusing instead on how to enumerate analysis decisions and display the results from the entire multiverse.
We visualize model fit in two ways: 
we use color to encode the NRMSE from a $k$-fold cross validation in the outcome view, and use predictive checks to compare observed data with model predictions in the model fit view.
Together these views show that a cross-product of analytic decisions can produce many universes with poor model fits, and many of the results that prior studies include in their overviews may not provide a sound base for subsequent inferences.
The prevalence of fit issues, which are immediately apparent in the Boba Visualizer, calls into question the idea that a multiverse analysis should consist of a cross-product of all \textit{a-priori} ``reasonable'' decisions.

We propose adding a step to the multiverse workflow where analysts must distinguish between what seems reasonable \textit{a-priori} vs. \textit{post-hoc}.
Boba supports this step in two ways: in the inference view we can use model averaging to produce a weighted combination of universes based on model fit, or we can simply omit universes below a certain model fit threshold chosen by the users. 
The latter option relies on analysts making a post-hoc subjective decision and might be susceptible to p-hacking.
However, one can pre-register a model quality threshold to eliminate this flexibility.
Should we enable more elaborate and interactive ways to give users control over pruning?
If so, how do we prevent analysts from unintentionally biasing the results?
These questions remain future work.

Indeed, a core tension in multiverse analysis is balancing the imperative of transparency with the \textit{need for principled reduction of uncertainty}.
Prior work on researcher degrees of freedom in analysis workflows~\cite{kale2019} identifies strategies that analysts use to make decisions (see also~\cite{lipshitz1997,boukhelifa2017}), including two which are relevant here:
\textit{reducing} uncertainty in the analysis process by following systematic procedures, and
\textit{suppressing} uncertainty by arbitrarily limiting the space of possible analysis paths.
In the context of Boba, design choices which direct the user's attention toward important information (e.g., highlighting models with good fit and decisions with a large influence on outcomes) and guide the user toward best practices (e.g., visual predictive checks) serve to push the user toward reducing rather than suppressing uncertainty.
Allowing users to interact with results as individual dots in the outcome view while showing aggregated uncertainty in the background reduces the amount of information that the user needs to engage with in order to begin exploring universes, while also maintaining a sense of the range of possible outcomes.
We believe that guiding users' attention and workflow based on statistical principles is critical.

\section{Conclusion}

This paper presents Boba, an integrated DSL and visual analysis system for authoring and interpreting multiverse analyses.
With the DSL, users annotate their analysis script to insert local variations, from which the compiler synthesizes executable script variants corresponding to all compatible analysis paths.
We provide a command line tool for compiling the DSL specification, running the generated scripts, merging the outputs, and invoking the visual analysis system.
We contribute a visual analysis system with linked views between analytic decisions and model estimates to facilitate systematic exploration of how decisions impact robustness, along with views for sampling uncertainty and model fit.
We also provide facilities for principled pruning of ``unreasonable'' specifications, and support inference to assess effect reliability.
Using Boba, we replicate two existing multiverse studies, gain a rich understanding of how decisions affect results, and find issues around uncertainty and model fit that change what we can reasonably take away from these multiverses.
Boba is available as open source software at \url{https://github.com/uwdata/boba}.

\acknowledgments{
We thank the anonymous reviewers, UW IDL members, Uri Simonsohn, Mike Merrill, Ge Zhang, Pierre Dragicevic, Yvonne Jansen, Matthew Kay, Brian Hall, Abhraneel Sarma, Fanny Chevalier, and Michael Moon for their help.
This work was supported by NSF Award 1901386.}

\bibliographystyle{abbrv-doi}

\bibliography{9-multiverse}
\end{document}